# MODÈLE D'ÉTALEMENT D'UNE GOUTTE EN PROJECTION PLASMA : APPLICATION AU REVETEMENT DE MATERIAUX THERMIQUEMENT DEGRADABLES

Khalid **FATAOUI,** Bernard **PATEYRON**[*], Nicolas **CALVÉ**, Mohammed **EL GANAOUI**

*SPCTS UMR6638 CNRS Université de Limoges*
*Faculté des Sciences et techniques 123 avenue Albert Thomas 87060 Limoges cedex*

## RÉSUMÉ

L'industrie de la projection plasma est confrontée à des besoins croissants de performance et à des exigences de plus en plus sévères. La réponse à ces contraintes impose une meilleure compréhension des phénomènes impliqués dans le processus et particulièrement l'impact des particules sur le substrat. En effet, dans le cas de substrats thermiquement dégradable comme le bois ou le béton, il s'agit de s'assurer que les températures à l'interface goutte-substrat restent acceptables.

Les travaux expérimentaux montrent que le phénomène d'étalement est contrôlé par les paramètres de la particule (état de fusion, température, vitesse, oxydation, …) et les paramètres du substrat (température, rugosité, état d'oxydation …). Lorsqu'une particule fondue entre en contact avec le substrat, son énergie cinétique est dissipée sous l'effet des forces de viscosité et de tension de surface, elle forme alors une lamelle. Cette phase d'étalement régit la qualité finale du dépôt (propriétés thermomécaniques) et la formation plus ou moins importante de porosité.

Nous proposons ici un modèle 2D axisymétrique couplant la vitesse d'impact et la solidification et permettant d'obtenir une corrélation du coefficient d'étalement dans le cas de matériaux tant céramiques que métalliques.

Mots clés : Ecrasement, impact, goutte, vitesse, dépôts

## NOMENCLATURE
**Symboles :**
*Lettres latines :*
$C_P$ capacité calorifique, $J.kg^{-1}.K^{-1}$
g accélération de la pesanteur, (9.81) $m.s^{-2}$
L chaleur latente, $J.kg^{-1}$
K conductivité thermique, $W.m^{-1}.K^{-1}$
$R_{th}$ résistance thermique de contact ($m^2.K/J$)
T température, $(K)$
V vitesse, $m.s^{-1}$
f

*Lettres grecques :*
$\Phi$ fonction de phase
$\sigma$ tension superficielle, $N.m^{-1}$
µ viscosité dynamique, $Pa.s$
ρ masse volumique, $kg.m^{-3}$
δ fonction de Dirac

*Indices / Exposants :*
f front de changement de phase
s solide
l liquid

## 1. INTRODUCTION

L'industrie de la projection plasma qui apporte des solutions à de nombreux défis industriels quant à l'énergétique et aux cycles de vie est sollicitée dans des applications extrêmement diversifiées telles que la déposition sur des substrats thermiquement fragiles. En effet, l'importance économique de procédés capables de réaliser par projection thermique des dépôts sur des matériaux aussi fragile que polymères organiques, bois ou bétons, est croissante.

Des gouttelettes liquides (voir **Figure 1** ) à température de fusion des céramiques (alumine, zircone) impactent ces surfaces à des vitesses qui peuvent être supersoniques [1, 2, 3]. Il s'agit de prédire par le calcul les températures atteintes à l'interface gouttelette-substrat lors de la cinétique d'étalement (voir **Figure 2** ). De telles prédictions ont été tentées par des modèles simplifiés qui donnent globalement satisfaction [4, 5, 6] eu égard à la rapidité de ces calculs.

Le modèle proposé est fondé sur les équations de Navier-Stokes, pour décrire l'écoulement et une équation supplémentaire prend en compte la tension superficielle et explique l'angle de contact. L'écoulement du fluide est supposé newtonien, et celui de la goutte est considéré incompressible. La seule tension s'appliquant à la surface libre est considérée comme normale à celle-ci. Les équations sont discrétisées par éléments finis.

La déformation libre de l'interface est suivie par une fonction de niveau $\phi$ utilisée pour déterminer la densité et la viscosité de part et d'autre de la surface adjacente séparant la gouttelette ($\phi = 1$) et l'atmosphère environnante ($\phi = 0$).

Les données des particules à l'impact (température et des vitesses pour un diamètre donné) sont issues tant du modèle de Jets&Poudres [7, 8, 9] pour des conditions usuelles de projection de particules de zircone en jet de plasma que de visualisation expérimentales [ 10, 11].

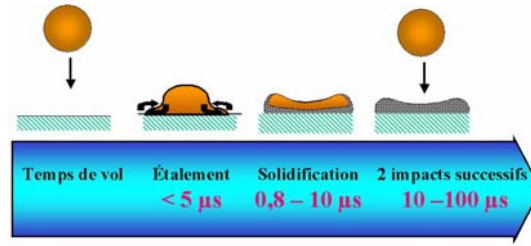

Figure 1 - Différentes échelles de temps lors de la formation du dépôt par projection [12].

## 2. MODÈLE DYNAMIQUE

Par application du principe fondamental de la dynamique, le bilan des forces au sein de chaque fluide newtonien exprime la vitesse *u* en fonction des paramètres du système :

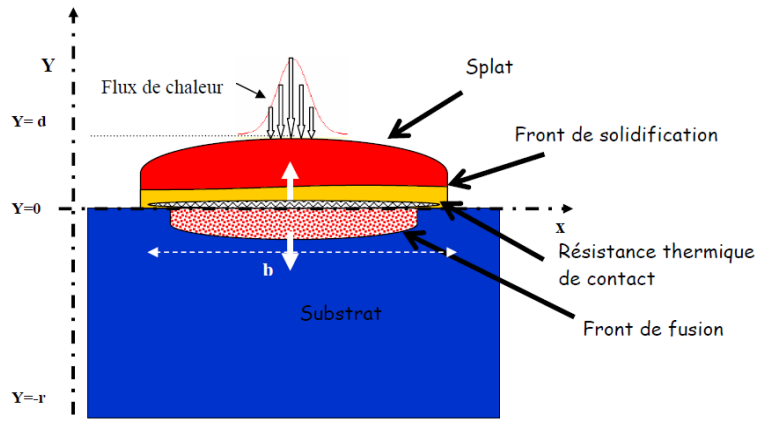

Figure 2 - Schéma de principe des échanges thermiques entre la goutte, en cours de gel, et le substrat selon [13].

$$\rho\left(\frac{\partial u}{\partial t} + \nabla.(uu)\right) = -\nabla p + \nabla(\mu \nabla u + \nabla u^T) + F_{TS} \quad (1)$$

$$\nabla u = 0$$

où ρ et μ sont respectivement la masse volumique et la viscosité du fluide, g l'accélération de la pesanteur. Le terme $F_{TS}$ représente les forces de tension de surface, il est lié au concept d'interface entre deux fluides non miscible. Cette force est représentée par

$$F_{Ts} = \sigma.k.\delta.n_i \quad (2)$$

où σ, k et δ sont respectivement le coefficient de tension superficielle, la courbure moyenne locale de l'interface et la fonction de Dirac.

Le modèle choisi ici est du type modèle à un fluide. En effet ce modèle s'adapte bien à la discrétisation sur maillage fixe cartésien. Ce type de modèle a été utilisé par plusieurs auteurs pour simuler la déposition sur substrat sec [14,15] ou l'impact sur film liquide [16]. Il convient de définir une fonction qui détermine géométriquement la position des phases dans le domaine. Soit la fonction de phase ϕ définie par (ϕ= 1 dans la zone contenant le fluide 1, ϕ = 0 dans la zone contenant le fluide 2 et *0< ϕ < 1* dans la zone d'interface entre les deux fluides.

La fonction de phase ϕ se déplace en fonction de la vitesse de l'interface *u*. Cette évolution est décrite par l'équation cinématique qui relie la fonction à la vitesse de l'interface *u* :

$$\frac{\partial \Phi}{\partial t} + u \nabla \Phi = 0 \quad (3)$$

Une propriété physique *λ* des milieux s'exprime en fonction de la présence de chaque fluide par :

$$\lambda = \lambda_1 + (\lambda_2 - \lambda_1)\phi \quad (4)$$

## 3. MODÈLE DE CHANGEMENT DE PHASE

Le milieu qui change de phase est représenté comme deux domaines séparés par une interface qui est le front de solidification. L'évolution de température de chaque milieu est résolue séparément, l'interface correspond à une condition dite de Stefan, définie par la conservation des flux à l'interface :

$$k_s \nabla T_s n - k_1 \nabla T_l n = \rho L_f V_f n \quad (5)$$

*l* et *s* représentent respectivement les phases liquide et solide et $V_f$ est la vitesse du front de solidification. La difficulté de cette méthode réside dans la nécessité de déterminer la position du front de solidification, ce qui est souvent effectué, soit par la méthode de Landau sur un maillage évolutif, soit sur maillage fixe en prenant en compte implicitement la condition de Stefan (méthode enthalpique) [7].

## 4. VALIDATION DU MODÈLE À DIFFÉRENTES VITESSES D'IMPACT

Il n'existe pas de formulation analytique qui permette de décrire avec précision l'évolution de l'impact et de l'étalement d'une goutte, pour valider la simulation. Nous avons donc confronté les résultats obtenus avec ceux de la référence [8]. En outre nous avons utilisé la formule analytique [9] afin de comparer le degré d'étalement maximal obtenu $\xi_{max}$, défini comme le rapport entre le diamètre maximal de la gouttelette étalée et le diamètre initial de la goutte.

$$\xi_{max} = \sqrt{\frac{w_e + 12}{3(1-\cos\theta) + 4(w_e/\sqrt{R_e})}} \quad (6)$$

avec $W_e$ et $R_e$ respectivement les nombres de Weber et de Reynolds d'impact, et $\theta$ l'angle de contact statique.

Considérons en coordonnées cartésiennes 2D axisymétrique l'impact d'une goutte sur un substrat. L'axe de la goutte est une condition limite de type symétrie afin de réduire le domaine. La condition aux limites à l'interface goutte-substrat est une condition de non glissement (**Figure 1**).

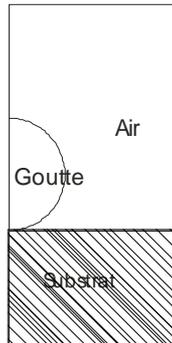

Figure 3 - Géométrie du système étudié : écrasement de la goutte sur la cible.

La simulation numérique est réalisée pour l'impact d'une goutte d'indium de 2mm de diamètre, de vitesse initiale de 0,1 m.s$^{-1}$ et dont les caractéristiques sont résumées dans [17,18].

Le pas de temps utilisé est constant $\Delta t = 5.10^{-6}$s. Les équations de Navier-Stokes en écoulement incompressible sont résolues par la méthode de Lagrange, et l'équation d'advection par la méthode '*level set*'.

Dans la simulation présentée, ici, le degré d'étalement maximum de la goutte est de l'ordre de $\xi = 2.11$. Dans la simulation [19], le degré d'étalement maximum de la goutte atteint une valeur de $\xi = 2.17$, soit un écart relatif de 2.7%, alors que l'expression analytique propose un degré d'étalement de $\xi = 2.07$, ce qui donne donc un écart relatif de 1.9% (Pasandideh-Fard et al. [14] proposent une dispersion des résultats inférieure à 15%).

La comparaison de deux modèle a été faite aussi par la fonction de phase $\phi$ (l'interface entre la goutte liquide et l'air), la **Figure 2** montre les résultats des deux modèles à l'instant t = 0.01s. Pour les faibles vitesses, les résultats sont donc cohérents avec ceux numérique de référence [19] et des expressions analytiques [20].

Pour les grandes vitesses (>100 m/s), nous avons comparé les résultats du modèle proposé avec ceux obtenus par Mauro Bertagnolli et al. [21]. La simulation numérique réalisée correspond à l'impact d'une particule de zircone de *20 μm* de diamètre et de température initiale *3400 K*, et animée d'une vitesse de *180 m.s$^{-1}$* à l'impact. Les caractéristiques de cette goutte sont données [18].

La **Figure 3**, compare l'étalement de la goutte de zircone dans le modèle proposé et dans celui de Mauro Bertagnolli [21]. L'étalement et la forme de la goutte sont semblables au cours de l'impact. À l'instant t = 0.5 μs, la goutte se stabilise et prend la forme de l'étalement final.

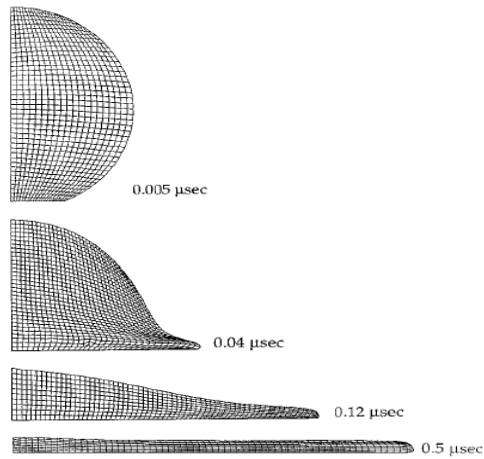

Figure 4 - Étalements avec maillage de la goutte de zircone fondue de 50 µm de diamètre selon Mauro Bertagnolli [21].

Les paramètres utilisés dans cette validation sont ceux traités par Madejski [22], Mauro Bertagnolli [21] et Yoshida [20]. Le modèle de Madejski est fondé sur l'hypothèse d'une déformation cylindrique de la goutte, le degré d'étalement maximum (équation 6) est alors estimé par :

$$\xi = MR_e^{0.2} = M\left(\frac{\rho V d}{\mu}\right)^{0.2} \qquad (7)$$

Selon Madejski, M est un coefficient constant égal à 1.2941 [8] alors que selon Yoshida et al. [10], M est égal à 0.925. Le modèle de Madejski surévalue légèrement le degré d'étalement relativement à celui de Yoshida. Les résultats du modèle présenté sont très proches de ceux obtenus par Mauro Bertagnolli [21]. Sur la **Figure 5** sont représentés les taux d'étalement prédits par ces différents modèles. Dans le diagramme logarithmique, la bande située entre les modèles Yoshida et de Madejski regroupe tous les résultats. En conséquence les résultats de la simulation de l'impact d'une goutte sur un substrat à une grande vitesse, sont donc cohérents avec les calculs théoriques et avec les autres modèles.

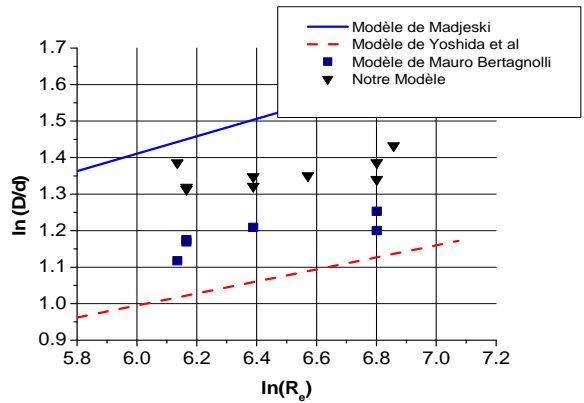

Figure 5 - Degré d'étalement en fonction du nombre de Reynolds.

## 5. EXPLOITATION DU MODÈLE

En projection plasma, le comportement de particules fondues à l'impact est un mécanisme élémentaire pour comprendre la microstructure de la couche résultante. En effet, les couches sont créées par l'empilement de lamelles formées par l'étalement et la solidification de gouttelettes individuelles fondues. Les particules projetées par plasma d'arc sont généralement de taille micrométrique (entre *10* et *100 µm*) en projection conventionnelle et leur vitesse d'impact s'étend de *50* à *350 m/s*.

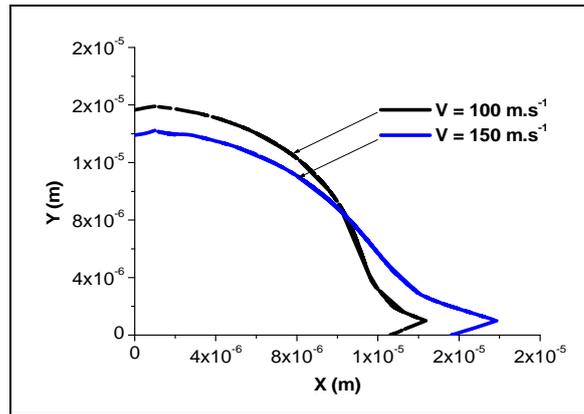

Figure 6 - Étalement de la goutte de zircone fondue de 50 µm de diamètre à l'instant 0.05 µs.

Cette étude est consacrée à une goutte de zircone préchauffée à *3500 K*, de *20 µm* de diamètre, projetée pour deux vitesses de *100 m.s$^{-1}$* et *150 m.s$^{-1}$* sur un substrat d'acier préchauffé à *700 K*.

Les **Figure 5** et **Figure 6** montrent différents instants de la modélisation de l'étalement d'une goutte de zircone de 20 µm sur le substrat avec plusieurs vitesses d'impact (100 et 150 m.s$^{-1}$). La goutte s'étale d'autant plus que la vitesse est plus élevée. Elle s'étale et se solidifie sur le substrat d'acier dans un temps inférieur à *1 µs*. A l'instant t = 0.3 µs, des éclaboussures apparaissent à partir de la goutte animée d'une vitesse de 150m/s, la rupture du fluide apparaît avec la naissance de gouttelettes secondaires éjectées parallèlement à la surface du substrat.

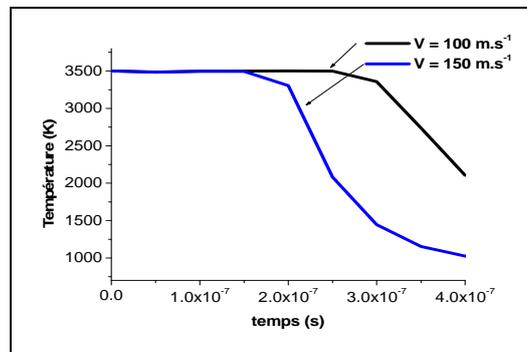

Figure 7 - Évolution de la température au point X=0 Y=50 µm *d'une goutte de zircone fondue frappant la cible à 100 m/s et 160 m/s.*

Sur la **Figure 7**, il apparaît que la goutte de zircone se solidifie d'autant plus vite que la vitesse d'impact est plus grande. Cela semble imputable au temps d'étalement plus rapide qui augmente donc les échanges thermiques entre la goutte et le substrat. Ainsi l'augmentation de la vitesse d'impact influe directement sur le temps d'étalement de la goutte et sur le diamètre maximum atteint par la lamelle.

Lorsque le liquide s'écoule sur la surface, il tend à suivre la forme de celle-ci. Plus cette dernière sera rugueuse plus le liquide aura des difficultés à remplir ses creux. Ainsi le liquide qui s'écoule perpendiculairement à ces obstacles est projeté verticalement par inertie.

Aux premiers instants qui suivent le contact de la goutte avec le substrat, la surface de contact et donc d'échange augmente rapidement et la résistance thermique de contact $R_{th}$ diminue de *5.10$^{-7}$* à *3.10$^{-7}$ m$^2$K/W* (**Figure 8**). Après étalement partiel, la surface de contact entre le substrat et la goutte augmente, alors que la température de la goutte reste quasi-constante. A ce moment, la température du substrat cesse de croître alors que la température de surface de la goutte s'abaisse. La résistance thermique de contact correspondante augmente pour atteindre un maximum à la fin de l'étalement. Ajoutons que pendant l'impact la goutte s'écrase sur la surface du substrat en générant une forte pression. La goutte liquide remplit toutes les cavités en surface du substrat, la résistance thermique de contact est minimale jusqu'à ce que la pression retombe, et alors la résistance thermique de contact augmente.

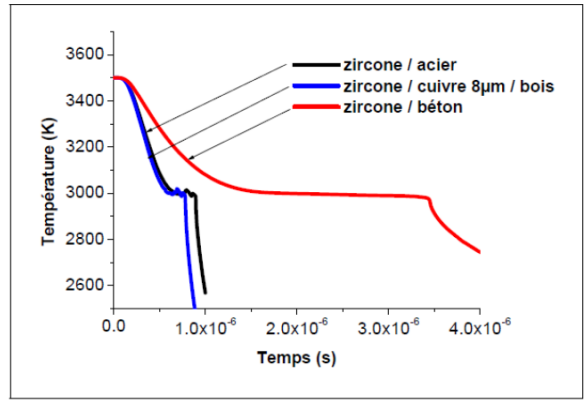

Figure 8 - Vitesse de solidification d'une goutte de zircone de 50 µm sur différents substrats dont acier béton et bois protégé par une sous couche de cuivre de 8µm.

Le calcul montre qu'une gouttelette de zircone de 50 µm à température de fusion 3500 K frappant différents support à la vitesse de 150 m/s gèle en des temps extrêmement différents (voir **Figure 8**). Il apparaît ainsi qu'une couche de protection du bois par 8 µm de cuivre est extrêmement efficace quant à la dissipation thermique de l'ensemble qui est alors comparable à un substrat d'acier.

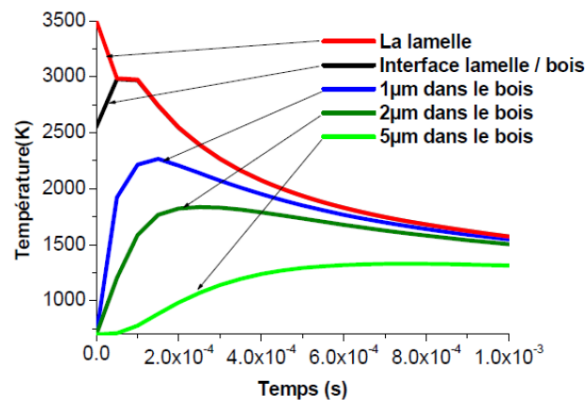

Figure 9 - Évolution de la température à différentes profondeur dans le support bois.

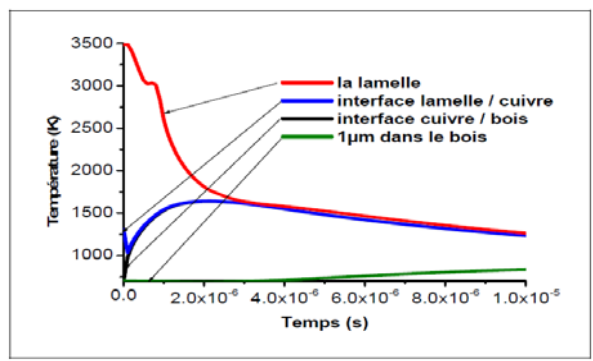

Figure 10 - Évolution temporelle de la température après impact d'une goutte de zircone de 50 µm sur un support bois protégé d'une lame de 3µm de cuivre.

La Figure 9 illustre la pénétration thermique en fonction du temps pour la même goutte de zircone fondue frappant une plaque de bois nu. Des températures de plus de 1200 K sont atteintes à une profondeur de 5 µm. Dans ces conditions, le bois est dégradé. La **Figure 10** montre par comparaison l'efficacité d'une couche de protection de 1 µm de cuivre. A 1 µm de profondeur, la température ne dépasse pas 600 K. Le bois est protégé de toute dégradation thermique.

# 6. CONCLUSION

Le modèle 2D de l'impact de gouttelettes permet donc d'estimer de façon efficace les grandeurs temporelles et dimensionnelles relatives aux phénomènes d'écrasement de gouttelette en cours de projection plasma.

La solidification est prise en compte lorsque la goutte atteint une température inférieure à la température de fusion ce qui contribue à figer plus rapidement la matière et diminue l'énergie cinétique du système ce qui diminue la vitesse d'étalement. De nombreux auteurs supposent que la gouttelette s'écrase en surfusion. Il serait donc intéressant de poursuivre l'étude en tenant compte de la surfusion afin de la représenter correctement et de l'intégrer dans le code global. Les perspectives immédiates de ce travail sont relatives à la mise en place de la simulation 3D de l'impact de gouttes où les transferts sont couplés aux changements de phase et à l'exploitation des calculs pour la recherche de condition de projection thermique sur des matériaux thermiquement dégradables.

## REMERCIEMENTS

## RÉFÉRENCES